\documentclass[12pt]{article}
\usepackage{graphicx}
\usepackage{cite}
\usepackage{amsmath}
\usepackage{float}
\def\beq{\begin{equation}}
\def\eeq{\end{equation}}
\def\bea{\begin{eqnarray}}
\def\eea{\end{eqnarray}}
\def\nn{\nonumber}

\def\roughly#1{\mathrel{\raise.3ex\hbox
{$#1$\kern-.75em\lower1ex\hbox{$\sim$}}}}
\def\lsim{\roughly<}

\def\s{\sqrt{2}}

\def\bs{B^0_s}
\def\bsbar{{\bar B}^0_s}

\def\bs{B^0_s}
\def\bsbar{{\bar B}^0_s}
\def\bsee{b \to s e^+ e^-}
\def\bsmumu{b \to s \mu^+ \mu^-}

\def\bsee{b \to s e^+ e^-}
\def\bsll{b \to s \ell^+ \ell^-}
\def\bsnunubar{b \to s \nu {\bar\nu}}

\def \cB{{\cal B}}
\def \SM{{\rm SM}}
\def \NP{{\rm NP}}

\begin{document}

\begin{flushright}
UdeM-GPP-TH-19-270 \\
\end{flushright}

\begin{center}
\bigskip
{\Large \bf \boldmath The $B$ Anomalies and New Physics in $\bsee$} \\
\bigskip
\bigskip
{\large
Alakabha Datta $^{a,}$\footnote{datta@phy.olemiss.edu},
Jacky Kumar $^{b,}$\footnote{jacky.kumar@umontreal.ca}
and David London $^{b,}$\footnote{london@lps.umontreal.ca}
}
\end{center}

\begin{flushleft}
~~~~~~~~~~~~~~~~~$a$: {\it Department of Physics and Astronomy, 108 Lewis Hall,}\\
~~~~~~~~~~~~~~~~~~~~~{\it University of Mississippi, Oxford, MS 38677-1848, USA}\\
~~~~~~~~~~~~~~~~~$b$: {\it Physique des Particules, Universit\'e de Montr\'eal,}\\
~~~~~~~~~~~~~~~~~~~~~{\it C.P. 6128, succ. centre-ville, Montr\'eal, QC, Canada H3C 3J7}
\end{flushleft}

\begin{center}
\bigskip (\today)
\vskip0.5cm {\Large Abstract\\} \vskip3truemm
\parbox[t]{\textwidth}{ We investigate the implications of the latest
  LHCb measurement of $R_K$ for NP explanations of the $B$ anomalies.
  The previous data could be explained if the $\bsmumu$ NP is in (I)
  $C_{9,\NP}^{\mu\mu}$ or (II) $C_{9,\NP}^{\mu\mu} =
  -C_{10,\NP}^{\mu\mu}$, with scenario (I) providing a better
  explanation than scenario (II). This continues to hold with the new
  measurement of $R_K$. However, for both scenarios, this measurement
  leads to a slight tension of $O(1\sigma)$ between separate fits to
  the $\bsmumu$ and $R_{K^{(*)}}$ data. In this paper, we investigate
  whether this tension can be alleviated with the addition of NP in
  $\bsee$. In particular, we examine the effect of adding such NP to
  scenarios (I) and (II). We find several scenarios in which this
  leads to improvements in the fits. $Z'$ and LQ models with
  contributions to both $\bsmumu$ and $\bsee$ can reproduce the data,
  but only within scenarios based on (II). If the tension persists in
  future measurements, it may be necessary to consider NP models with
  more than one particle contributing to $\bsll$.}

\end{center}


\thispagestyle{empty}
\newpage
\setcounter{page}{1}
\baselineskip=14pt

At present, there are several measurements of $B$-decay processes
involving the transition $\bsll$ ($\ell = \mu,e$) that are in
disagreement with the predictions of the standard model (SM). First,
there are discrepancies with the SM in a number of observables in $B
\to K^* \mu^+\mu^-$ \cite{BK*mumuLHCb1, BK*mumuLHCb2, BK*mumuBelle,
  BK*mumuATLAS, BK*mumuCMS} and $B_s^0 \to \phi \mu^+ \mu^-$
\cite{BsphimumuLHCb1, BsphimumuLHCb2}, decays which involve only
$\bsmumu$. Second, the measurements of $R_K \equiv \cB(B^+ \to K^+
\mu^+ \mu^-)/{\cal B}(B^+ \to K^+ e^+ e^-)$ \cite{RKexpt} and $R_{K^*}
\equiv {\cal B}(B^0 \to K^{*0} \mu^+ \mu^-)/\cB(B^0 \to K^{*0} e^+
e^-)$ \cite{RK*expt} also disagree with the SM predictions. These
ratios involve both $\bsmumu$ and $\bsee$. In this paper, we refer to
these two sets of observables as the $\bsmumu$ and $R_{K^{(*)}}$
observables.

Since all processes involve $\bsmumu$, it is natural to examine
whether the $B$ anomalies can be explained by adding new physics (NP)
to this decay.  The $\bsmumu$ transitions are defined via an effective
Hamiltonian with vector and axial vector operators:
\bea
H_{\rm eff} &=& - \frac{\alpha G_F}{\s \pi} V_{tb} V_{ts}^*
      \sum_{a = 9,10} ( C_a O_a + C'_a O'_a ) ~, \nn\\
O_{9(10)} &=& [ {\bar s} \gamma_\mu P_L b ] [ {\bar\mu} \gamma^\mu (\gamma_5) \mu ] ~,
\label{Heff}
\eea
where the $V_{ij}$ are elements of the Cabibbo-Kobayashi-Maskawa (CKM)
matrix and the primed operators are obtained by replacing $L$ with
$R$. The Wilson coefficients (WCs) include both the SM and NP
contributions: $C_X = C_{X,\SM} + C_{X,\NP}$. Following the
announcement of the $R_{K^*}$ measurement in 2017, global fits were
performed that combine the various $\bsll$ observables
\cite{Capdevila:2017bsm, Altmannshofer:2017yso, DAmico:2017mtc,
  Hiller:2017bzc, Geng:2017svp, Ciuchini:2017mik, Celis:2017doq,
  Alok:2017sui}. It was found that the net discrepancy with the SM is
at the level of 4-6$\sigma$, and that the data can be explained if the
nonzero WCs are (I) $C_{9,\NP}^{\mu\mu}$ or (II) $C_{9,\NP}^{\mu\mu} =
-C_{10,\NP}^{\mu\mu}$. In Ref.~\cite{Alok:2017sui}, the best-fit
values of the WCs for these two scenarios were found to be (I)
$C_{9,\NP}^{\mu\mu} = -1.20 \pm 0.20$ and (II) $C_{9,\NP}^{\mu\mu} =
-C_{10,\NP}^{\mu\mu} = -0.62 \pm 0.14$ (other analyses found similar
results). The simplest NP models involve the tree-level exchange of a
leptoquark (LQ) or a $Z'$ boson. Scenario (II) can arise in LQ or $Z'$
models, but scenario (I) is only possible with a $Z'$
\cite{Alok:2017sui}.

The first measurement of $R_K$ was made in 2014 by the LHCb
Collaboration using the Run 1 data \cite{RKexpt}. For $1.1 \le q^2 \le
6.0 ~{\rm GeV}^2$, where $q^2$ is the dilepton invariant mass-squared,
the result was
\beq
R_{K,{\rm Run~1}}^{\rm old} = 0.745^{+0.090}_{-0.074}~{\rm (stat)} \pm 0.036~{\rm (syst)} ~.
\label{RKold}
\eeq
This differs from the SM prediction of $R_K^\SM = 1 \pm 0.01$
\cite{IsidoriRK} by $\sim 2.6\sigma$.  Recently, LHCb announced new
$R_K$ results \cite{LHCbRKnew}. First, the Run I data was reanalyzed
using a new reconstruction selection method. The new result is
\beq
R_{K,{\rm Run~1}}^{\rm new} = 0.717^{+0.083}_{-0.071}~{\rm (stat)} ^{+0.017}_{-0.016}~{\rm (syst)} ~.
\eeq
Second, the Run 2 data was analyzed:
\beq
R_{K,{\rm Run~2}} = 0.928^{+0.089}_{-0.076}~{\rm (stat)} \pm ^{+0.020}_{-0.017}~{\rm (syst)} ~.
\eeq
Combining the Run 1 and Run 2 results, the LHCb measurement of $R_K$
is
\beq
R_K = 0.846^{+0.060}_{-0.054}~{\rm (stat)} ^{+0.016}_{-0.014}~{\rm (syst)} ~.
\label{RKmeas}
\eeq
This is closer to the SM prediction, though the discrepancy is still
$\sim 2.5\sigma$ due to the smaller errors. 

The LHCb measurement of $R_{K^*}$ was \cite{RK*expt}
\beq
R_{K^*} = 
\begin{cases}
0.660^{+0.110}_{-0.070}~{\rm (stat)} \pm 0.024~{\rm (syst)} ~,~~ 0.045 \le q^2 \le 1.1 ~{\rm GeV}^2 ~, \\
0.685^{+0.113}_{-0.069}~{\rm (stat)} \pm 0.047~{\rm (syst)} ~,~~ 1.1 \le q^2 \le 6.0 ~{\rm GeV}^2 ~.
\end{cases}
\label{RK*meas}
\eeq
Recently, Belle announced its measurement of $R_{K^*}$
\cite{BelleRK*new}:
\beq
R_{K^*} =
\begin{cases}
0.52^{+0.36}_{-0.26} \pm 0.05 ~,~~ 0.045 \le q^2 \le 1.1 ~{\rm GeV}^2 ~, \\
0.96^{+0.45}_{-0.29} \pm 0.11 ~,~~ 1.1 \le q^2 \le 6.0 ~{\rm GeV}^2 ~, \\
0.90^{+0.27}_{-0.21} \pm 0.10 ~,~~ 0.1 \le q^2 \le 8.0 ~{\rm GeV}^2 ~, \\
1.18^{+0.52}_{-0.32} \pm 0.10 ~,~~ 15.0 \le q^2 \le 19.0 ~{\rm GeV}^2 ~, \\
0.94^{+0.17}_{-0.14} \pm 0.08 ~,~~ 0.045 \le q^2 ~.
\end{cases}
\eeq
The errors are considerably larger than in the LHCb measurement.

In this paper we examine the effect of these new measurements --
especially that of $R_K$ [Eq.~(\ref{RKmeas})] -- on the NP
explanations of the $\bsll$ $B$ anomalies.

The first step is to simply combine all the observables, and update
the global fit performed in Ref.~\cite{Alok:2017sui}. (We refer to
this paper for a description and the measured values of all the
(CP-conserving) $\bsmumu$ observables.) This is done using the
programs {\tt MINUIT} \cite{James:1975dr,James:2004xla,James:1994vla},
{\tt flavio} \cite{Straub:2018kue} and {\tt Wilson}
\cite{Aebischer:2018bkb}. The results are shown in Table
\ref{globalfit}. 

\begin{table}[h]
\begin{center}
\begin{tabular}{|c|c|c|c|} \hline
Scenario & WC & p-value & pull  \\
\hline
(I) $C^{\mu\mu}_{9,\NP} $ & $ -1.10 \pm 0.16 $ & 0.71 & 5.8  \\
\hline
(II) $C_{9,\NP}^{\mu\mu} = -C_{10,\NP}^{\mu\mu} $ & $ -0.53 \pm 0.08$  & 0.64 & 5.5  \\
\hline
\end{tabular}
\end{center}
\caption{Best-fit values of the WCs (taken to be real), the p-value,
  and the pull = $\sqrt{\chi^2_{\SM} - \chi^2_{\SM + \NP}}$ for the
  global fit including all $\bsmumu$ and $R_{K^{(*)}}$ observables.
  For each case there are 115 degrees of freedom.
\label{globalfit}}
\end{table}

For each scenario we present the best-fit value of the WCs, as well as
the p-value and the pull:
\begin{enumerate}

\item The p-value is derived from $\chi^2_{\rm min}/{\rm d.o.f.}$ and
  characterizes the goodness of fit. If all observables were
  ``clean,'' i.e., if the theoretical error associated with their
  predictions were small, then the dominant error in the fit would be
  purely statistical.  In this case, the $\chi^2_{\rm min}/{\rm
    d.o.f.}$ distribution would be Gaussian, with a central value of
  1, corresponding to a p-value of 0.5. In general, it is assumed
  that, if the fit produces a p-value of $< 0.05$ (i.e., outside the
  95\% C.L.\ region), this is considered to be an unacceptable fit.

  Usually, one does not compare the p-values of different fits -- a
  fit is either acceptable or it is not. However, in this paper, we
  are interested in determining whether a particular (acceptable)
  scenario provides a better description of the data than another
  (acceptable) scenario, and so we will compare the p-values.
  (Admittedly, the difference in the p-values of two acceptable
  scenarios is not statistically significant.)

  In the present fit, the $\bsmumu$ observables are not clean: all of
  them involve sizeable theoretical uncertainties (form factors), and
  each analysis of the $B$ anomalies has its own method of treating
  these theoretical errors. (In this paper, we take the theoretical
  uncertainties into account following Ref.~\cite{Straub:2015ica}.)
  However, the point is that the way these theoretical errors are
  estimated affects the results of the fit: methods with large (small)
  theoretical errors will tend to have larger (smaller) p-values.
  Thus, it makes no sense to compare the p-values of analyses that use
  different methods of dealing with the theoretical uncertainties.  On
  the other hand, what {\it is} rigorous is to compare the p-values of
  scenarios that use the same theoretical method.  We therefore
  conclude that scenario (I) (p-value = 0.71) provides a slightly
  better explanation of the data than scenario (II) (p-value = 0.64).
  And both are enormous improvements on the SM, which has a p-value of
  0.05.

\item The pull is defined to be $\sqrt{\chi^2_{\SM} - \chi^2_{\SM +
    \NP}}$, i.e., it quantifies how much better the SM + NP fit is
  than the fit with the SM alone. In the present case, since both
  scenarios involve only one free parameter, a pull of 5.8 indicates
  that (i) the discrepancy between the experimental data and the
  predictions of the SM is at least $5.8\sigma$, and (ii) the addition
  of NP improves the agreement with the measurements by
  $5.8\sigma$. From the p-values, we already concluded that scenario
  (I) explains the data somewhat better than scenario (II); in Table
  \ref{globalfit}, this is reflected in a larger pull. Of course, this
  does not exclude the possibility of finding an even larger
  improvement over the SM in another NP scenario.

\end{enumerate}

While this is an interesting result, the global fit does not contain
all the important NP implications of the experimental data. Let us
instead separate the data into $\bsmumu$ and $R_{K^{(*)}}$
observables, and perform separate fits on these two data sets. The
results are shown in Table \ref{separatefits}. We see that there is
now a slight tension between the NP WCs required to explain the
$\bsmumu$ and $R_{K^{(*)}}$ data: in scenario (I), the two best-fit
values differ by $1.1\sigma$, while in scenario (II) the difference is
$1.3\sigma$, where $\sigma$ is defined by adding the errors of the two
solutions in quadrature. The most obvious explanation of this tension
is that it is simply a statistical fluctuation. However, in this
paper, we investigate whether the tension can be alleviated with the
addition of NP in $\bsee$. With this in mind, we consider a variety of
scenarios in which some NP $\bsee$ WCs are taken to be nonzero, in
order to see if this tension can be removed, and the fit improved. As
we will see, there are a number of scenarios with NP in $\bsee$ in
which this occurs.

\begin{table}[htb]
\begin{center}
\begin{tabular}{|c|c|c|} \hline
Scenario & Data Set & WC \\
\hline
(I) $C_{9,\NP}^{\mu\mu} $ & $R_{K^{(*)}}$ & $ -0.82 \pm 0.28 $  \\
& $\bsmumu$ & $ -1.17 \pm 0.18 $  \\
\hline
(II) $C_{9,\NP}^{\mu\mu}  = -C_{10,\NP}^{\mu\mu} $ & $R_{K^{(*)}}$ & $ -0.38 \pm 0.11 $  \\
& $\bsmumu$ & $ -0.62 \pm 0.14 $  \\
\hline
\end{tabular}
\end{center}
\caption{Best-fit values of the WCs (taken to be real) for separate
  fits including the $\bsmumu$ or $R_{K^{(*)}}$ observables.
\label{separatefits}}
\end{table}

In a recent paper \cite{Alguero:2019pjc}, a similar observation was
made about the different NP implications of the $\bsmumu$ and
$R_{K^{(*)}}$ data. And in Ref.~\cite{Alguero:2018nvb}, it was argued
that a better description of the data can be obtained if one adds NP
to the NP already assumed to be present in $\bsmumu$ WCs. However, in
both Refs.~\cite{Alguero:2019pjc} and \cite{Alguero:2018nvb}, rather
than focusing on additional NP in $\bsmumu$ and/or $\bsee$, there the
analysis is done in terms of lepton-flavour-universal (LFU) and
lepton-flavour-universality-violating (LFUV) NP. This same type of
language is used in Ref.~\cite{StraubMoriondtalk}. There it is argued
that, when one includes the latest $R_K$ and $R_{K^*}$ measurements in
the fit, a better description of the data is obtained if one has
additional LFU NP. One of the points of the present paper is to note
that this is not the only possibility. Here we show that additional NP
in $\bsee$, which is clearly LFUV NP, can also lead to a better
description of the data.

We begin by investigating the addition of NP in $\bsee$ to scenario
(I). We examine three different scenarios, shown in Table
\ref{NPbseescenariosI}. In scenario S0, the best-fit value of the
$\bsee$ WC is consistent with zero, as in Table \ref{globalfit}. This
is reflected in the fact that the pull is also unchanged from Table
\ref{globalfit}. Thus, S0 is no better than the original scenario (I),
and we discard it. On the other hand, in scenarios S1 and S2, nonzero
values of the $\bsee$ WCs are preferred. Furthermore, these scenarios
{\it are} clear improvements, as is indicated by the increased
p-values and pulls. These scenarios demonstrate that, by adding NP to
$\bsee$, one can improve the agreement with the data.

\begin{table*}[htb] 
\renewcommand{\arraystretch}{2}
\begin{center}
\begin{tabular}{|c|c|c|c|c|}
 \hline 
& $C_{9,\NP}^{\mu\mu}$ & NP in $\bsee$ & p-value & Pull \\
\hline
S0 & $-1.04 \pm 0.19$ & $C_{9,\NP}^{ee} = -0.09 \pm 0.33$ & 0.73 & 5.8 \\
\hline
S1 & $-1.03 \pm 0.18$ & $C_{9,\NP}^{\prime ee} = -0.41 \pm 0.28$ & 0.77 & 6.0 \\
\hline
S2 & $-1.12 \pm 0.17$ & $C_{10,\NP}^{\prime ee} = 0.42 \pm 0.25$ & 0.76 & 5.9 \\
\hline
\end{tabular}
\end{center}
\caption{Scenario (I) with the addition of one nonzero NP WC in
  $\bsee$: best-fit values of the WCs (taken to be real), the p-value,
  and the pull.}
\label{NPbseescenariosI}
\end{table*}

\begin{figure}[H]
\begin{center}
\includegraphics[width=0.49\textwidth]{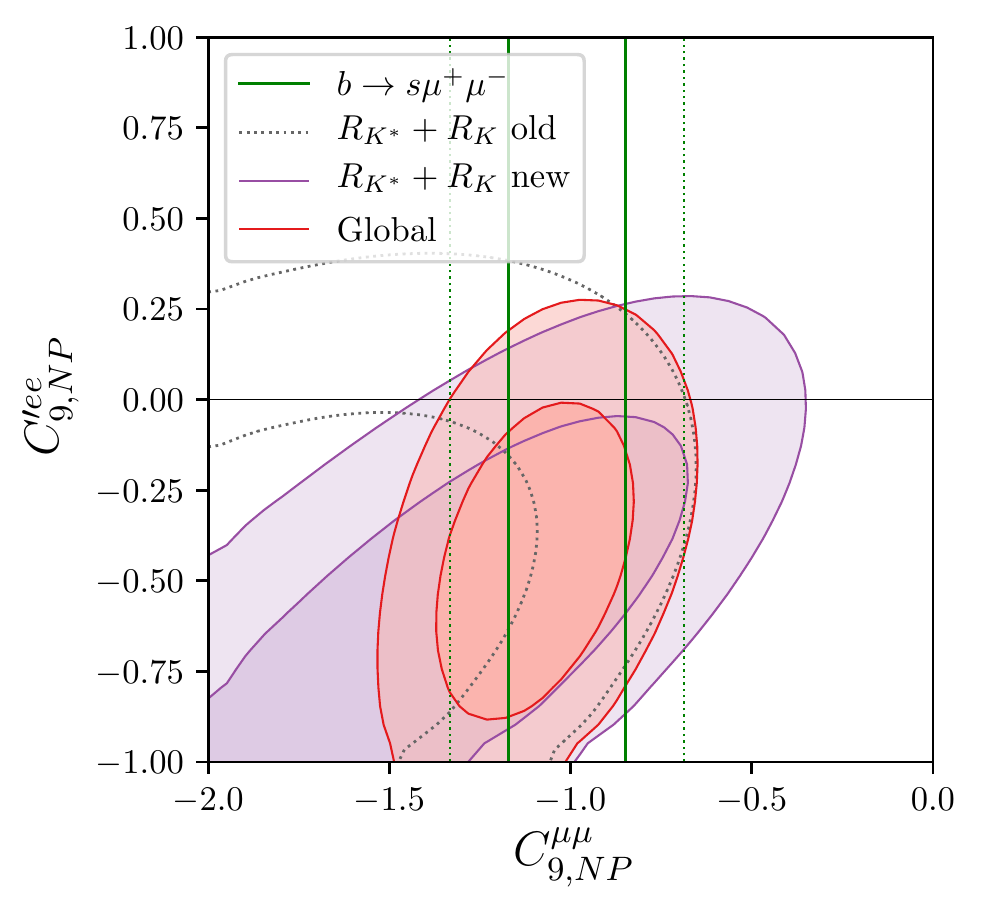}
\includegraphics[width=0.49\textwidth]{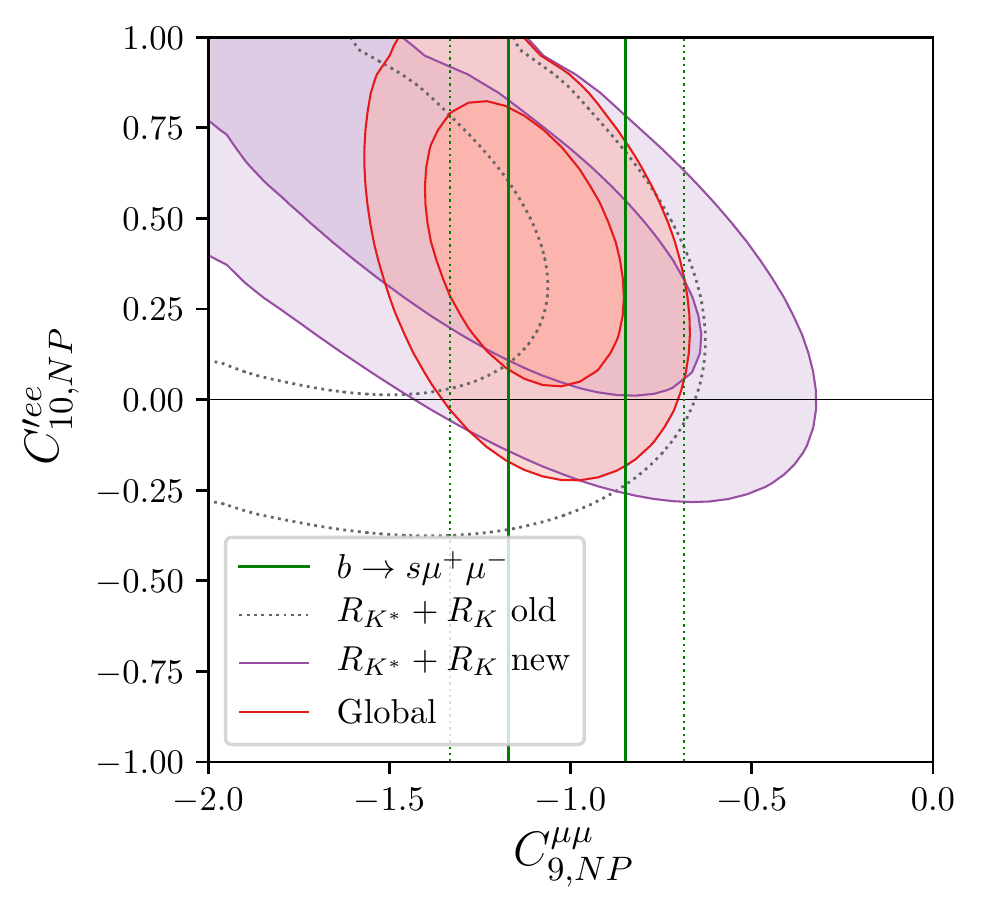}
\caption{Scenarios S1 (left) and S2 (right): the allowed $1\sigma$ and
  $2\sigma$ regions of the $\bsmumu$ (vertical lines) and new
  $R_{K^{(*)}}$ (mauve) observables, as well as the combined global
  fit (red), are shown as functions of $C_{9,\NP}^{\mu\mu}$ and
  $C_{9,\NP}^{\prime ee}$ (left) or $C_{10,\NP}^{\prime ee}$
  (right). The $1\sigma$ and $2\sigma$ regions associated with the old
  $R_{K^{(*)}}$ observables are indicated by the dotted lines.}
\label{S1S2fits}
\end{center}
\end{figure}

For scenarios S1 and S2, in Fig.~\ref{S1S2fits} we show the allowed
$1\sigma$ and $2\sigma$ regions of the $\bsmumu$ and new $R_{K^{(*)}}$
observables individually, as well as the combined fit, all as
functions of the WCs. In both cases, we see that the combined global
fit prefers nonzero values of the $\bsee$ WC. We also see how the new
measurement of $R_K$ has moved the parameter space of the combined
fit.

We now add NP in $\bsee$ to scenario (II).  The three different
scenarios considered are shown in Table \ref{NPbseescenariosII}. In
all cases, there is an improvement in the fits compared to Table
\ref{globalfit}.

\begin{table*}[htb] 
\renewcommand{\arraystretch}{2}
\begin{center}
\begin{tabular}{|c|c|l|c|c|}
 \hline 
& $C_{9,\NP}^{\mu\mu} = -C_{10,\NP}^{\mu\mu}$ & NP in $\bsee$ & p-value & Pull \\
\hline
S3 & $-0.67 \pm 0.15$ & $C_{9,\NP}^{ee} = -C_{10,\NP}^{ee} = -0.28 \pm 0.20$ & 0.65 & 5.6 \\
\hline
S4 & $-0.64 \pm 0.14$ & $C_{9,\NP}^{ee} = -0.65 \pm 0.44$ & 0.69 & 5.7 \\
\hline
S5 & $-0.56 \pm 0.09$ & $C_{9,\NP}^{\prime ee} = -C_{10,\NP}^{\prime ee} = -0.25 \pm 0.14$ & 0.67 & 5.6 \\
\hline
\end{tabular}
\end{center}
\caption{Scenario (II) with the addition of one nonzero NP WC in
  $\bsee$: best-fit values of the WCs (taken to be real), the p-value,
  and the pull.}
\label{NPbseescenariosII}
\end{table*}

\begin{figure}[H]
\begin{center}
\includegraphics[width=0.44\textwidth]{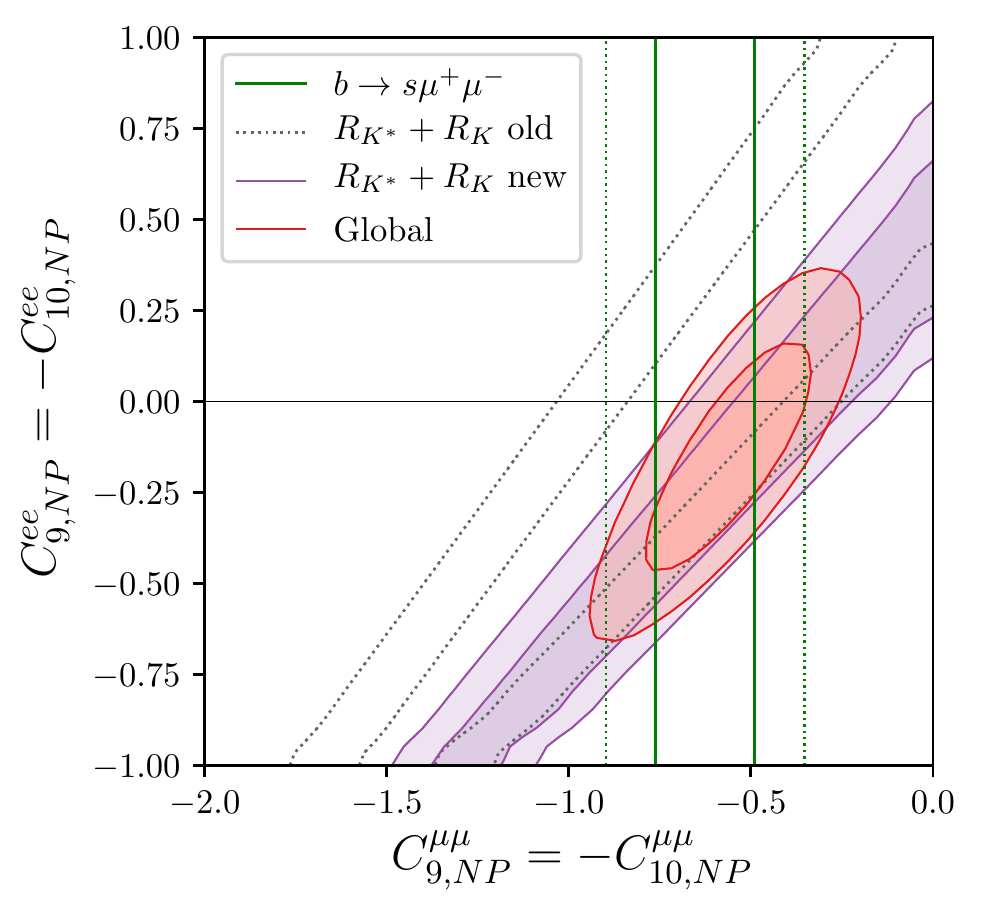} \\
\includegraphics[width=0.44\textwidth]{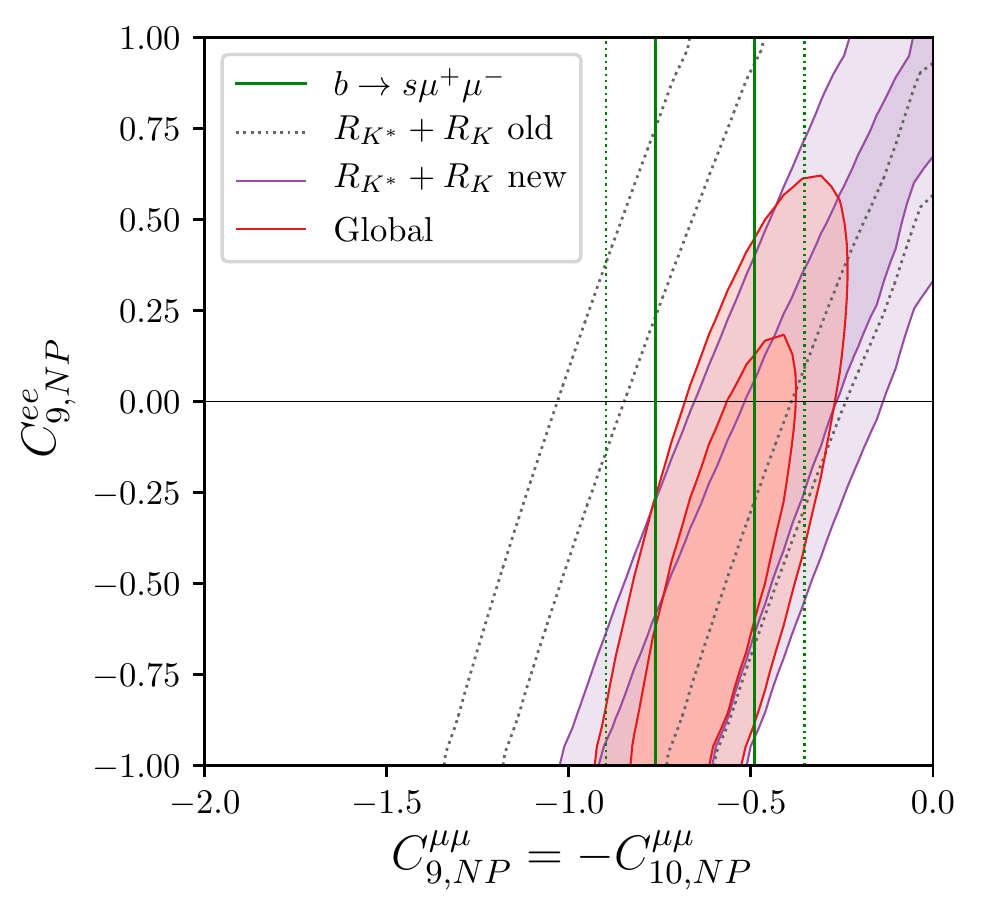} 
\includegraphics[width=0.44\textwidth]{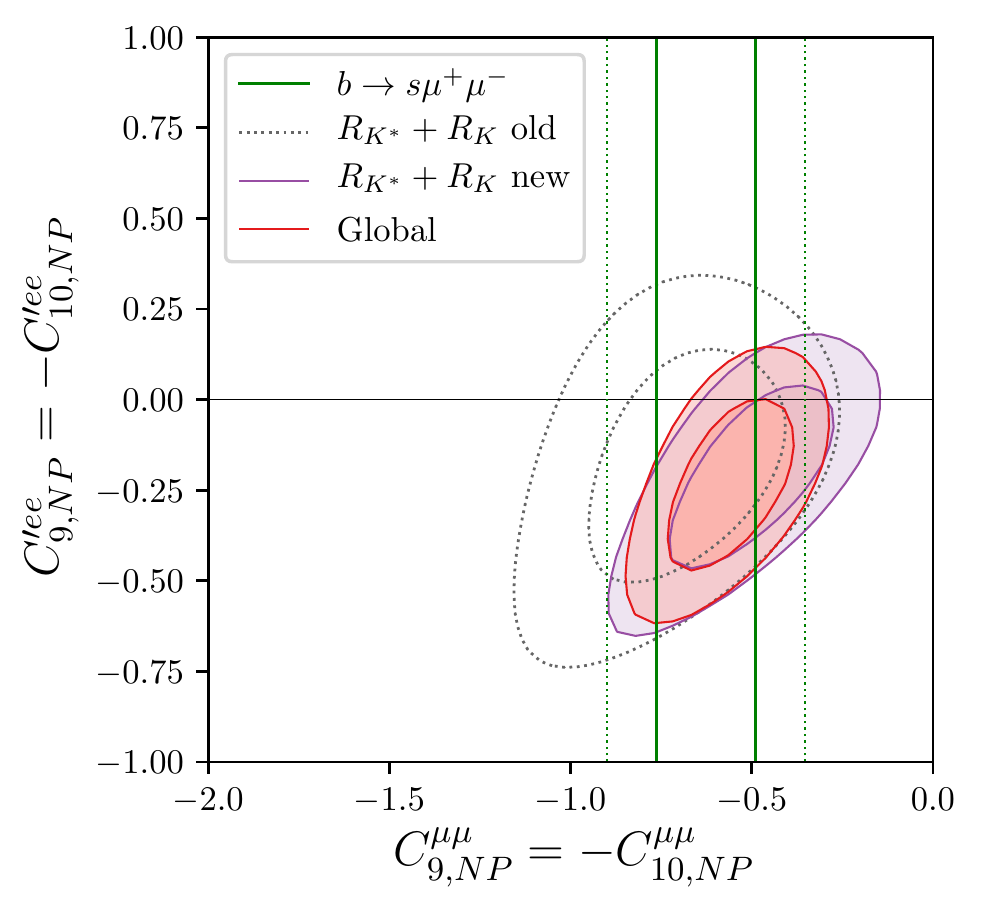}
\caption{Scenarios S3 (upper), S4 (lower left) and S5 (lower right):
  the allowed $1\sigma$ and $2\sigma$ regions of the $\bsmumu$
  (vertical lines) and new $R_{K^{(*)}}$ (mauve) observables, as well
  as the combined global fit (red), are shown as functions of the
  WCs. The $1\sigma$ and $2\sigma$ regions associated with the old
  $R_{K^{(*)}}$ observables are indicated by the dotted lines.}
\label{S3S4S5fits}
\end{center}
\end{figure}

We therefore see that, with the addition of NP in $\bsee$, scenarios
S3-S5 show an improvement over scenario (II) of Table \ref{globalfit}.
Still, even in the best case (S4), where the p-value and pull increase
to 0.69 and 5.7, respectively, one still does not quite reach the
level of scenario (I) without the addition of NP in $\bsee$ (Table
\ref{globalfit}).  That is, even if we allow for NP in $\bsee$,
scenario (I) continues to provide a better explanation of the data
than scenario (II).  Even so, solutions S3-S5 are in no way ruled out,
and so should not be discarded. In Fig.~\ref{S3S4S5fits} we show the
allowed regions of the S3, S4 and S5 scenarios in the parameter space
of the WCs.

We now turn to a model-dependent analysis. As noted earlier, the
simplest NP models that contribute to $\bsll$ involve the tree-level
exchange of a $Z'$ boson [scenario (I) or (II)] or a LQ [scenario (II)
  only]. With the previous data, both of these NP models were viable.
Does this still hold with the present data? We begin by looking at
LQs.

There are three types of LQ that can contribute to $\bsll$ at tree
level and involve only left-handed particles ($C_{9,\NP}^{\mu\mu} =
-C_{10,\NP}^{\mu\mu}$). They are an $SU(2)_L$-triplet scalar ($S_3$),
an $SU(2)_L$-singlet vector ($U_1$), and an $SU(2)_L$-triplet vector
($U_3$) \cite{Sakakietal}. Their couplings are
\bea
{\cal L}_{S_3} & = & y'_{\ell q}\, {\bar\ell}^c_L i \tau_2 {\vec \tau} q_L \cdot {\vec S}_3 + h.c., \nn\\
{\cal L}_{U_1} & = & (g_{\ell q}\, {\bar\ell}_L \gamma_\mu q_L + g_{ed}\, {\bar e}_R \gamma_\mu d_R) U_1^\mu + h.c., \nn\\
{\cal L}_{U_3} & = & g'_{\ell q}\, {\bar\ell}_L \gamma_\mu {\vec \tau} q_L \cdot {\vec U}_3^\mu + h.c.
\label{LQlist}
\eea
Here, in the fermion currents and in the subscripts of the couplings,
$q$ and $\ell$ represent left-handed quark and lepton $SU(2)_L$
doublets, respectively, while $u$, $d$ and $e$ represent right-handed
up-type quark, down-type quark and charged lepton $SU(2)_L$ singlets,
respectively. The LQs can couple to fermions of any generation. To
specify which particular fermions are involved, we add superscripts to
the couplings. For example, $g^{\prime \mu s}_{\ell q}$ is the
coupling of the $U_3$ LQ to a left-handed $\mu$ (or $\nu_\mu$) and a
left-handed $s$ (or $c$).  These couplings are relevant for $\bsmumu$
or $\bsee$ (and possibly $\bsnunubar$).

In LQ models, there may be contributions to lepton-flavour-conserving
operators in addition to $O^{(\prime)\ell\ell}_{9,10}$ ($\ell =
e,\mu$) [Eq.~(\ref{Heff})]. They are
\bea
& O^{(\prime)\ell\ell}_\nu = [ {\bar s} \gamma_\mu P_{L(R)} b ] [ {\bar\nu}_\ell \gamma^\mu (1 - \gamma_5) \nu_\ell ] ~, & \nn\\
& O^{(\prime)\ell\ell}_S = [ {\bar s} P_{R(L)} b ] [ {\bar\ell} \ell ] ~~,~~~~
O^{(\prime)\ell\ell}_P = [ {\bar s} P_{R(L)} b ] [ {\bar\ell} \gamma_5 \ell ] & ~.
\label{newops}
\eea
$O^{(\prime)\ell\ell}_\nu$ contributes to $b \to s \nu_\ell {\bar
  \nu}_\ell$, while $O^{(\prime)\ell\ell}_S$ and
$O^{(\prime)\ell\ell}_P$ are additional contributions to $b \to s
\ell^+ \ell^-$. There may also be contributions to the
lepton-flavour-violating (LFV) operators
\bea
& O^{(\prime)\ell\ell'}_{9(10)} = [ {\bar s} \gamma_\mu P_{L(R)} b ] [ {\bar\ell} \gamma^\mu (\gamma_5) \ell' ] ~, & \nn\\
& O^{(\prime)\ell\ell'}_\nu = [ {\bar s} \gamma_\mu P_{L(R)} b ] [ {\bar\nu}_\ell \gamma^\mu (1 - \gamma_5) \nu_{\ell'} ] ~, & \nn\\
& O^{(\prime)\ell\ell'}_S = [ {\bar s} P_{R(L)} b ] [ {\bar\ell} \ell' ] ~~,~~~~
O^{(\prime)\ell\ell'}_P = [ {\bar s} P_{R(L)} b ] [ {\bar\ell} \gamma_5 \ell' ] & ~,
\label{newopsLFV}
\eea
where $\ell,\ell' = e,\mu$, with $\ell \ne \ell'$.
$O^{(\prime)\ell\ell'}_{9(10)}$, $O^{(\prime)\ell\ell'}_S$ and
$O^{(\prime)\ell\ell'}_P$ contribute to $\bs \to e^\pm \mu^\mp$ and $B
\to K^{(*)} e^\pm \mu^\mp$. Using the couplings in Eq.~(\ref{LQlist}),
one can compute which WCs are affected by each LQ. These are shown in
Table~\ref{LQWC} for $\ell = \ell' = \mu$ \cite{AGC}, and it is
straightforward to change one $\mu$ or both to an $e$. Finally, there
may also be a 1-loop contribution to the LFV decay $\mu \to e \gamma$:
\beq
O^{(L)R}_\gamma = [ {\bar e} \sigma_{\mu \nu} P_{L(R)} \mu ] F_{\mu\nu} ~.
\label{ogamma}
\eeq
All LFV operators can arise if there is a single LQ that couples to
both $\mu$ and $e$. Since the constraints from LFV processes are
extremely stringent, we therefore anticipate that it may be difficult
for a single LQ to both explain the $B$ anomalies via couplings to
$\bsmumu$ and $\bsee$ and satisfy the LFV constraints.

\begin{table}
\begin{center}
\begin{tabular}{|c|cccc|} \hline
LQ & $C_{9,\NP}^{\mu\mu}$ & $C_{10,\NP}^{\mu\mu}$ & $C_{9,\NP}^{\prime\mu\mu}$ & $C_{10,\NP}^{\prime\mu\mu}$ \\
   & $C_{S,\NP}^{\mu\mu}$ & $C_{S,\NP}^{\prime\mu\mu}$ & $C_{\nu,\NP}^{\mu\mu}$ & $C_{\nu,\NP}^{\prime\mu\mu}$ \\
\hline
$S_3$ & $y_{\ell q}^{\prime \mu b} (y_{\ell q}^{\prime \mu s})^*$
                             & $- y_{\ell q}^{\prime \mu b} (y_{\ell q}^{\prime \mu s})^*$ & 0 & 0 \\
& 0 & 0 & $\frac12 y_{\ell q}^{\prime \mu b} (y_{\ell q}^{\prime \mu s})^*$ & 0 \\
\hline
$U_1$ & $- g_{\ell q}^{\mu b} (g_{\ell q}^{\mu s})^*$ & $ g_{\ell q}^{\mu b} (g_{\ell q}^{\mu s})^*$
& $- g_{e d}^{\mu b} (g_{e d}^{\mu s})^*$ & $- g_{e d}^{\mu b} (g_{e d}^{\mu s})^*$ \\
& $2 g_{\ell q}^{\mu b} (g_{e d}^{\mu s})^*$ & $2 (g_{\ell q}^{\mu s})^* g_{e d}^{\mu b}$ & 0 & 0 \\
\hline
$U_3$ & $- g_{\ell q}^{\prime \mu b} (g_{\ell q}^{\prime \mu s})^*$
                                   & $ g_{\ell q}^{\prime \mu b} (g_{\ell q}^{\prime \mu s})^*$ & 0 & 0 \\
& 0 & 0 & $- 2 g_{\ell q}^{\prime \mu b} (g_{\ell q}^{\prime \mu s})^*$ & 0 \\
\hline
\end{tabular}
\end{center}
\caption{Contributions of the different LQs to the $\bsmumu$ WCs of
  various operators. The $\bsee$ WCs are obtained by changing $\mu \to
  e$ in the superscripts.  The normalization $K \equiv \pi / (\sqrt{2}
  \alpha G_F V_{tb} V_{ts}^* M_{LQ}^2)$ has been factored out. For
  $M_{LQ} = 1$ TeV, $K = -644.4$.
\label{LQWC}}
\end{table}

The analysis of the LQ models has the following ingredients:
\begin{itemize}

\item $\bsmumu$ and $\bsee$: All LQs have $C_{9,\NP}^{\ell\ell} = -
  C_{10,\NP}^{\ell\ell}$, $\ell=\mu,e$.  In principle, the $U_1$ LQ
  could also produce $C_{9,\NP}^{\prime\ell\ell} = +
  C_{10,\NP}^{\prime\ell\ell}$. However, if these primed WCs are
  sizeable, so too are the scalar WCs $C_{S,\NP}^{\ell\ell}$ and
  $C_{S,\NP}^{\prime\ell\ell}$ (see Table \ref{LQWC}). Now, the scalar
  operators $O^{(\prime)\ell\ell}_S$ [Eq.~(\ref{newops})] contribute
  significantly to $\bs\to\ell^+\ell^-$ \cite{Alok:2010zd}. The
  present measurement of $\cB(\bs\to\mu^+\mu^-)$ \cite{Aaij:2013aka,
    CMS:2014xfa}, in agreement with the SM, and the upper bound
  $\cB(\bs\to e^+e^-) < 2.8 \times 10^{-7}$ (90\% C.L.)
  \cite{pdg2018} constrain $C_{9,\NP}^{\prime\ell\ell} = +
  C_{10,\NP}^{\prime\ell\ell}$ to be small.

\item $b \to s \nu_\ell {\bar \nu}_{\ell^{(\prime)}}$: As can be seen
  in Table~\ref{LQWC}, the $S_3$ and $U_3$ LQs can have nonzero
  $C_{\nu,\NP}^{(\prime) \mu\mu}$ WCs, so there may be additional
  constraints from $b \to s \nu_\ell {\bar \nu}_{\ell^{(\prime)}}$.
  However, it was shown in Ref.~\cite{Alok:2017jgr} that the present
  constraints from $B \to K^{(*)} \nu {\bar\nu}$ are rather weak, and
  do not place significant limits on the WCs.

\item LFV processes: The contributions of LQs to LFV processes were
  examined in detail in Ref.~\cite{Kumar:2019qbv}. It was found that
  the most important LFV process is $\mu \to e \gamma$, with $\cB(\mu
  \to e \gamma) < 4.2 \times 10^{-13}$ (90\% C.L.) \cite{pdg2018}.
  Even though the LQ contributes only at the 1-loop level, the very
  small upper limit on the branching ratio places stringent
  constraints on the model. The relevant WCs are
  \cite{Crivellin:2017dsk}
\beq
C_{\gamma}^{L} = -\frac{e N_c m_\mu}{16 \pi^2  M_{LQ}^2} \,
\frac{n}{K} \left ( \xi C_{9,\NP}^{ee} + \frac{1}{\xi} C_{9,\NP}^{\mu \mu} \right ) ~~,~~~~ C_{\gamma}^R = 0 ~,
\eeq
where $\xi \equiv g_{\ell q}^{\mu b}/g_{\ell q}^{es}$, $K$ is given in
the caption of Table \ref{LQWC}, and $n= \frac{1}{8}$, 2 and
$\frac{1}{6}$ for the $S_3$, $U_3$ and $U_1$ LQ models,
respectively. In computing the constraints on the LQ models from $\mu
\to e \gamma$, we conservatively take $\xi=2$, as it leads to the
weakest constraints.

\end{itemize}

Given that LQs can only contribute to $C_{9,\NP}^{\ell\ell} = -
C_{10,\NP}^{\ell\ell}$, $\ell=\mu,e$, the only one of the scenarios in
Tables \ref{NPbseescenariosI} and \ref{NPbseescenariosII} that can be
generated by LQs is S3, which is based on scenario (II). Indeed, the
S3 and LQ fits are quite similar, except that there is an additional
constraint on LQ models from $\mu \to e \gamma$. We find that all
three LQ models can explain the data, with pulls of 5.6 ($U_1$), 5.5
($S_3$), and 5.5 ($U_3$). The pulls are very slightly lower than that
of S3, due to the additional $\mu \to e \gamma$ constraint.  We
therefore conclude that, with the new $R_K$ data, explanations of the
$B$ anomalies involving a single LQ with contributions to both
$\bsmumu$ and $\bsee$ are still possible, though they do not reproduce
the data quite as well as NP scenarios based on scenario (I) (i.e., S1
amd S2).

We now turn to $Z'$ models. As was the case for LQs, other processes
may be affected by $Z'$ exchange, and these produce constraints on the
couplings. In particular, the ${\bar s}b Z'$ coupling is constrained
by $\bs$-$\bsbar$ mixing and the $\mu^+ \mu^- Z'$ coupling is
constrained by the production of $\mu^+\mu^-$ pairs in
neutrino-nucleus scattering, $\nu_\mu N \to \nu_\mu N \mu^+ \mu^-$
(neutrino trident production). These constraints are discussed in
detail in Ref.~\cite{Kumar:2019qbv}. There it is found that, when
these constraints are taken into account, the expected sizes of the
$\bsmumu$ NP WCs are $|C_{9,10,\NP}^{(\prime)\mu\mu}| \lsim 0.6$.

In the most general case, the couplings of the $Z'$ to the various
pairs of fermions are independent. For $\bsmumu$ and $\bsee$
transitions, the couplings that interest us are $g_L^{sb}$,
$g_R^{sb}$, $g_L^\mu$, $g_R^\mu$, $g_L^e$ and $g_R^e$, which are the
coefficients of $({\bar s} \gamma^\mu P_L b)Z'_\mu$, $({\bar s}
\gamma^\mu P_R b)Z'_\mu$, $({\bar \mu} \gamma^\mu P_L \mu)Z'_\mu$,
$({\bar \mu} \gamma^\mu P_R \mu)Z'_\mu$, $({\bar e} \gamma^\mu P_L
e)Z'_\mu$ and $({\bar e} \gamma^\mu P_R e)Z'_\mu$,
respectively. Defining $g_V^\ell \equiv g_R^\ell + g_L^\ell$ and
$g_A^\ell \equiv g_R^\ell - g_L^\ell$ $(\ell = \mu, e)$, we can then
write
\bea
& C_{9,\NP}^{\mu\mu} = K \, g_L^{sb} g_V^\mu ~,~~
C_{10,\NP}^{\mu\mu} = K \, g_L^{sb} g_A^\mu ~,~~
C_{9,\NP}^{\prime\mu\mu} = K \, g_R^{sb} g_V^\mu ~,~~
C_{10,\NP}^{\prime\mu\mu} = K \, g_R^{sb} g_A^\mu ~, & \nn\\
& C_{9,\NP}^{ee} = K \, g_L^{sb} g_V^e ~,~~
C_{10,\NP}^{ee} = K \, g_L^{sb} g_A^e ~,~~
C_{9,\NP}^{\prime ee} = K \, g_R^{sb} g_V^e ~,~~
C_{10,\NP}^{\prime ee} = K \, g_R^{sb} g_A^e ~, &
\eea
where $K$ is given in the caption of Table \ref{LQWC}.

With these expressions, it is straightforward to see that scenarios
S1, S2 and S5 of Tables \ref{NPbseescenariosI} and
\ref{NPbseescenariosII} cannot be produced with a $Z'$. On the other
hand, scenarios S3 and S4 can (scenario S0 can as well, but it has
been discarded). Both scenarios require $g_L^{sb} \ne 0$ and $g_R^{sb}
= 0$, while scenario S3 (S4) requires $g_A^\mu = -g_V^\mu$ ($g_A^\mu =
0$). In addition, the WCs roughly satisfy
$|C_{9,10,\NP}^{(\prime)\mu\mu}| \lsim 0.6$, which is required by the
constraints from $\bs$-$\bsbar$ mixing and neutrino trident
production. This shows that scenarios S3 and S4 can be generated in a
model with a $Z'$ gauge boson. Still, S3 and S4 are part of scenario
(II), which does not explain the data quite as well as scenario (I).

To summarize, the NP models containing a single new particle that
contributes to $\bsll$ at tree level -- LQ models and models with a
$Z'$ -- can both explain the present data if there are contributions
to both $\bsmumu$ and $\bsee$. However, in both cases, the nonzero
$\bsmumu$ WCs are $C_{9,\NP}^{\mu\mu} = -C_{10,\NP}^{\mu\mu}$
[scenario (II)], and this does not provide quite as good a fit to the
data as those scenarios with only $C_{9,\NP}^{\mu\mu} \ne 0$. This
leads one to consider the possibility of more than one NP
contribution.  Indeed, realistic NP models often contain a variety of
new particles. To investigate the possibilities, it is useful to
approach this question from the SM Effective Field Theory (SMEFT)
\cite{Buchmuller:1985jz, Grzadkowski:2010es} point of view.

Any NP model must respect the $SU(3)_C \times SU(2)_L \times U(1)_Y$
gauge symmetries of the SM. When this NP is integrated out, one
produces operators involving only the SM particles, but these
must also be invariant under the SM symmetries. There are,
of course, many possible operators, but we are interested only in
those that contribute to the WCs $C_{9,10}^{(\prime)\ell\ell}$ ($\ell
= \mu$ or $e$) at low energy. Restricting ourselves to dimension-six
NP operators that contribute to $\bsll$ at tree level, there are two
categories. First, there are four-fermion operators:
\bea
{\cal O}_{\ell q}^{(1)} &=& ({\bar\ell}_i \gamma_\mu \ell_j) ({\bar q}_k \gamma^\mu q_l) ~, \nn\\
{\cal O}_{\ell q}^{(3)} &=& ({\bar\ell}_i \gamma_\mu \tau^I \ell_j) ({\bar q}_k \gamma^\mu \tau^I q_l) ~, \nn\\
{\cal O}_{q e} &=& ({\bar q}_i \gamma_\mu q_j) ({\bar e}_k \gamma^\mu e_l) ~, \nn\\
{\cal O}_{\ell d}^{(1)} &=& ({\bar\ell}_i \gamma_\mu \ell_j) ({\bar d}_k \gamma^\mu d_l) ~, \nn\\
{\cal O}_{e d}^{(1)} &=& ({\bar e}_i \gamma_\mu e_j) ({\bar d}_k \gamma^\mu d_l) ~.
\eea
Second, there are operators involving the Higgs field:
\bea
{\cal O}_{\varphi q}^{(1)} &=& (\varphi^\dagger i \overleftrightarrow{D}_\mu \varphi ) ({\bar q}_i \gamma^\mu q_j) ~, \nn\\
{\cal O}_{\varphi q}^{(3)} &=& (\varphi^\dagger i \overleftrightarrow{D}_\mu^I \varphi ) ({\bar q}_i \tau^I \gamma^\mu q_j) ~, \nn\\
{\cal O}_{\varphi d} &=& (\varphi^\dagger i \overleftrightarrow{D}_\mu \varphi ) ({\bar d}_i \gamma^\mu d_j) ~.
\eea

The $\bsll$ WCs can be written in terms of the coefficients of these
operators \cite{Aebischer:2015fzz}. The NP four-fermion operators
generate
\bea
C_{9,\NP}^{ij} &=& \frac{\pi}{\alpha} \frac{v^2}{\Lambda^2} 
\left[ {\tilde C}_{\ell q}^{(1)ij23} + {\tilde C}_{\ell q}^{(3)ij23} + {\tilde C}_{q e}^{23ij} \right] ~, \nn\\
C_{10,\NP}^{ij} &=& \frac{\pi}{\alpha} \frac{v^2}{\Lambda^2} 
\left[ {\tilde C}_{q e}^{23ij} - {\tilde C}_{\ell q}^{(1)ij23} - {\tilde C}_{\ell q}^{(3)ij23} \right] ~,
\eea
and
\bea
C_{9,\NP}^{\prime ij} &=& \frac{\pi}{\alpha} \frac{v^2}{\Lambda^2} 
\left[ {\tilde C}_{\ell d}^{ij23} + {\tilde C}_{e d}^{ij23} \right] ~, \nn\\
C_{10,\NP}^{\prime ij} &=& \frac{\pi}{\alpha} \frac{v^2}{\Lambda^2} 
\left[ {\tilde C}_{e d}^{ij23} - {\tilde C}_{\ell d}^{ij23} \right] ~.
\eea
The $ij = \mu\mu$ and $ee$ WCs are not necessarily equal, so these are
LFUV NP contributions. (These have been studied in
Ref.~\cite{Celis:2017doq}.) The operators involving the Higgs field
generate LFU NP contributions:
\bea 
C_{9,\NP}^{ii} &=& \frac{\pi}{\alpha} \frac{v^2}{\Lambda^2} 
\left[ {\tilde C}_{\varphi q}^{(1)23} + {\tilde C}_{\varphi q}^{(3)23} \right] (-1 + 4 \sin^2 \theta_W ) ~, \nn\\ 
C_{10,\NP}^{ii} &=& \frac{\pi}{\alpha} \frac{v^2}{\Lambda^2} 
\left[ {\tilde C}_{\varphi q}^{(1)23} + {\tilde C}_{\varphi q}^{(3)23} \right] ~, \nn\\ 
C_{9,\NP}^{\prime ii} &=& \frac{\pi}{\alpha} \frac{v^2}{\Lambda^2} {\tilde C}_{\varphi d}^{23} (-1 + 4 \sin^2 \theta_W ) ~, \nn\\ 
C_{10,\NP}^{\prime ii} &=& \frac{\pi}{\alpha} \frac{v^2}{\Lambda^2} {\tilde C}_{\varphi d}^{23} ~.  
\eea
The $C_{S,\NP}^{(\prime)}$ and $C_{P,\NP}^{(\prime)}$ WCs can be
treated similarly, but we note that they are not independent in SMEFT
\cite{AGC}.

Thus, if one wishes to generate a particular $\bsll$ WC, the above
indicates which NP operators are required. The last step is to
establish which types of NP particules can generate these NP
operators. This was examined in Ref.~\cite{deBlas:2017xtg}. In Table
\ref{table:SPI}, we present the list of all types of NP particles and
the operators that they generate. This allows model builders to work
out exactly $\bsll$ WCs are generated in a particular model.
Conversely, if one wishes to generate only a particular WC, one can
compute which combinations of particles are necessary to do this.

\begin{table*}[t] 
\renewcommand{\arraystretch}{2}
\begin{center}
\begin{tabular}{|c|c|c|c|}
\hline
Field & Operators & Field &  Operators \\ 
\hline
Coloured Spin-0 & $\mathcal{O}_{\ell q}^{(1)}, \mathcal{O}_{\ell q}^{(3)}$ & 
$SU(2)_L$-singlet  &$\mathcal{O}_{\ell q}^{(1)}, \mathcal{O}_{\ell q}^{(3)}, \mathcal{O}_{qe}, \mathcal{O}_{\ell d}$, \\
$SU(2)_L$-singlet & & 
Vector Boson & $\mathcal{O}_{ed}, \mathcal{O}_{\phi q}^{(1)}, \mathcal{O}_{\phi d}$ \\ 
\hline
Coloured Spin-0 &$\mathcal{O}_{qe}$ &
$SU(2)_L$-doublet  & $\mathcal{O}_{\phi q}^{(1)},\mathcal{O}_{\phi q}^{(3)} ,\mathcal{O}_{\phi d}$  \\
$SU(2)_L$-doublet & &
Vector Boson & \\ 
\hline
Coloured Spin-0 & $\mathcal{O}_{\ell q}^{(1)}, \mathcal{O}_{\ell q}^{(3)}$ &
$SU(2)_L$-triplet  & $\mathcal{O}_{\ell q}^{(3)}, \mathcal{O}_{\phi q}^{(3)}$ \\
$SU(2)_L$-triplet & &
Vector Boson & \\ 
\hline
Exotic Quark: &$\mathcal{O}_{\phi q}^{(1)}, \mathcal{O}_{\phi q}^{(3)}$ &
Coloured Spin-1 &  $\mathcal{O}_{\ell q}^{(1)}, \mathcal{O}_{\ell q}^{(3)}, \mathcal{O}_{ed}$\\
$SU(2)_L$ Vector-singlet  & &
$SU(2)_L$-singlet  & \\ 
\hline
Exotic Quark: &$\mathcal{O}_{\phi d}$ &
Coloured Spin-1 &  $\mathcal{O}_{qe}$\\
$SU(2)_L$ Vector-doublet & &
$SU(2)_L$-doublet  & \\ 
\hline
Exotic Quark: & $\mathcal{O}_{\phi q}^{(1)}, \mathcal{O}_{\phi q}^{(3)}$ & 
Coloured Spin-1 &  $\mathcal{O}_{\ell q}^{(1)}, \mathcal{O}_{\ell q}^{(3)}$\\ 
$SU(2)_L$ Vector-triplet  & &
$SU(2)_L$-triplet  & \\ 
\hline
\end{tabular}
\end{center}
\caption{One-particle extensions of SM that contribute to $b\to s\ell
  \ell$ at tree level \cite{deBlas:2017xtg}.}
\label{table:SPI}
\end{table*}

To conclude, in this paper we have examined the NP implications of the
latest measurements of $R_K$ and $R_{K^*}$. The $R_K$ result is
particularly important. There are two sets of observables: (i) those
involving only $\bsmumu$ decays, and (ii) $R_{K^{(*)}}$, which involve
both $\bsmumu$ and $\bsee$ transitions. If a global fit to all $\bsll$
data is performed, assuming new physics only in $\bsmumu$, it is found
that (i) there is still a sizeable (5-6$\sigma$) discrepancy between
the experimental results and the predictions of the SM, and (ii) this
type of NP can explain it. However, if one looks more closely and
performs separate fits to the $\bsmumu$ and $R_{K^{(*)}}$ data, there
is now a slight tension: the two fits give results that differ by
$O(1\sigma)$. This may well be simply a statistical fluctuation, but
here we examine whether the addition of NP in $\bsee$ can reduce the
tension.

It has been shown model-independently that the previous data could be
explained by the addition of NP in (I) $C_{9,\NP}^{\mu\mu}$ or (II)
$C_{9,\NP}^{\mu\mu} = -C_{10,\NP}^{\mu\mu}$, with scenario (I)
providing a better explanation than scenario (II). We considered the
addition of NP in $\bsee$ to these scenarios to see if the agreement
with the present data can be improved.  We identified several
scenarios in which the addition of nonzero $\bsee$ WCs to (I) or (II)
resulted in such improvements. It has been argued elsewhere
\cite{StraubMoriondtalk} that an improved agreement with the data can
be obtained if there is additional lepton-flavour-universal NP. Our
results show that this is not the only possibility: additional NP in
$\bsee$, which is clearly lepton-flavour-universality-violating NP,
can also do the job.

We also performed a model-dependent analysis. For NP models that
involve the tree-level exchange of a single particle (LQ models and
models with a $Z'$), we showed that they could explain the data, but
only within scenarios based on (II). Since scenarios based on (I)
provide a slightly better explanation of the data, it may be that more
than one NP particle is contributing to $\bsll$.  Using an SMEFT
approach, we identify which NP operators contribute to $\bsll$ at tree
level, and what types of NP particles lead to these operators. This
will permit the building of models that generate the desired $\bsmumu$
and $\bsee$ WCs.

\bigskip
\noindent
{\bf Acknowledgments}: This work was financially supported in part by
NSERC of Canada (JK, DL).

\end{document}